# Isoprene and acetone concentration profiles during exercise on an ergometer

J King[1,2,3], A Kupferthaler[1,2], K Unterkofler[3,2], H Koc[3,2,5], S Teschl[4],
G Teschl[5], W Miekisch[6,2], J Schubert[6,2], H Hinterhuber[7,2] and A Amann[1,2,*,#]

[1] Innsbruck Medical University, Department of Operative Medicine, Anichstr. 35, A-6020 Innsbruck, Austria

[2] Breath Research Unit of the Austrian Academy of Sciences, Dammstr. 22, A-6850 Dornbirn, Austria

[3] Vorarlberg University of Applied Sciences, Hochschulstr. 1, A-6850 Dornbirn, Austria

[4] University of Applied Sciences Technikum Wien, Höchstädtplatz 5, A-1200 Wien, Austria

[5] Universität Wien, Fakultät für Mathematik, Nordbergstr. 15, A-1090 Wien, Austria

[6] University of Rostock, Department of Anaesthesiology and Intensive Care, Schillingallee 35, D-18057 Rostock, Germany

[7] Innsbruck Medical University, Department of Psychiatry, Anichstr. 35, A-6020 Innsbruck, Austria

*Corresponding author:* Anton Amann, Univ.-Clinic for Anesthesia, Anichstr. 35, A-6020 Innsbruck, Austria, email: anton.amann@i-med.ac.at, anton.amann@oeaw.ac.at.

[#] Dr. Amann is representative of Ionimed GesmbH, Innsbruck

# Abstract

A real-time recording setup combining exhaled breath VOC measurements by proton transfer reaction mass spectrometry (PTR-MS) with hemodynamic and respiratory data is presented. Continuous automatic sampling of exhaled breath is implemented on the basis of measured respiratory flow: a flow-controlled shutter mechanism guarantees that only end-tidal exhalation segments are drawn into the mass spectrometer for analysis.

Exhaled breath concentration profiles of two prototypic compounds, isoprene and acetone, during several exercise regimes were acquired, reaffirming and complementing earlier experimental findings regarding the dynamic response of these compounds reported by Senthilmohan et al. [1] and Karl et al. [2]. While isoprene tends to react very sensitively to changes in pulmonary ventilation and perfusion due to its lipophilic behavior and low Henry constant, hydrophilic acetone shows a rather stable behavior. Characteristic (median) values for breath isoprene concentration and molar flow, i.e., the amount of isoprene exhaled per minute are 100 ppb and 29 nmol/min, respectively, with some intra-individual day-to-day variation. At the onset of exercise breath isoprene concentration increases drastically, usually by a factor of ~ 3-4 within about one minute. Due to a simultaneous increase in ventilation, the associated rise in molar flow is even more pronounced, leading to a ratio between peak molar flow and molar flow at rest of ~ 11.

Our setup holds great potential in capturing continuous dynamics of non-polar, low-soluble VOCs over a wide measurement range with simultaneous appraisal of decisive physiological factors affecting exhalation kinetics. In particular, data appear to favor the hypothesis that short-term effects visible in breath isoprene levels are mainly caused by changes in pulmonary gas exchange patterns rather than fluctuations in endogenous synthesis.

# Introduction

The basic requirement in *real-time* breath gas analysis is to develop an experimental setup enabling the fast quantification of volatile organic compounds (VOCs) in exhaled breath as well as for the acquisition of additional physiological variables in a synchronized, reproducible and non-invasive way. The first part of this contribution provides an extensive proposal on how to achieve this aim and documents the necessary features. Adjoined to this technical report, the second part outlines a series of *real-time* exhaled breath measurements during specified ergometer workload sequences which were carried out with the purpose of reaffirming and complementing earlier experimental findings in this area, e.g., reported by Senthilmohan et al. [1] and Karl et al. [2].

Regarding the analysis of exhaled breath samples, we limit ourselves to Proton-Transfer-Reaction Mass Spectrometry (PTR-MS), which is a relatively new analytical technique for determining concentration levels of volatile molecular species down to ppt level on the basis of chemical ionization [3, 4]. Due to its high sensitivity for a large variety of trace gases commonly occurring in human breath, PTR-MS has proven to be a valuable and rapid quantification tool in breath-related VOC research [5, 6]. Depending on the number of different mass-to-charge ratios considered, data acquisition can be performed on a time scale of approximately one second, which theoretically offers the possibility of drawing and analyzing several breath samples per exhalation cycle. Hence, contrary to other analytical methods such as gas chromatography mass spectrometry (GC-MS), which often require time-consuming preconcentration steps [7-11], PTR-MS can be used for measurement of continuous VOC profiles in real time. This is particularly important for detection of metabolic effects manifesting themselves in short-lived changes of certain marker compound levels [12-15]. It is advisable to complement such profiles with additional physiological data influencing VOC concentrations: cardiac output, which controls the rate at which trace gases circulate from organs and periphery to the lung, or alveolar ventilation, which governs the transport of VOCs through the respiratory tree. Combining the real-time capability of PTR-MS with systems recording hemodynamic and/or respiratory factors remains an ambitious task from an experimental point of view. The two main difficulties in the development of a reliable and robust *real-time* measurement tool are:

   (a) a consistent integration of all sensor devices guaranteeing synchronized data gathering
   (b) the standardization of the breath sampling procedure itself.

While (a) is mainly an issue regarding adequate instrument and software engineering, aim (b) is at the heart of exhaled breath analysis and still a matter of ongoing debate [16]. A major concern here is to ensure the extraction of end-tidal air, which can be implemented by flow- or $CO_2$-controlled sampling in order to selectively detect different exhalation phases. Some approaches have been suggested to prevent exhaled breath samples from being diluted with fresh dead space air [17-19]: nevertheless, most of the real-time experiments published in literature use mixed-expiratory breath samples. The same holds true for off-line breath sampling, which traditionally is carried out by collecting and storing the breath sample in some container prior to analysis (e.g., Tedlar bags).

Motivated by this lack of a standardized breath sampling procedure, a breath sampling device (not shown) has been developed in our laboratory which serves to fill Tedlar bags with alveolar air in a flow and/or $CO_2$-controlled manner, taking into account the onset of the alveolar plateau [20] during exhalation phases. Here we present a generalization of this method for automatic *real-time* sampling. In addition, we describe a recording setup efficiently combining hemodynamic as well as respiratory and VOC-related data streams and monitoring of these variables during ergometer-induced workload scenarios.

**Materials and Methods**

An experimental setup capable of measuring the multitude of physiological signals mentioned above consists of five central parts:

(A) a hemodynamical monitor measuring heart rate, heart minute volume, blood pressure, etc.
(B) a spirometer measuring the volumetric flow rate when breathing through a flow transducer of some form (e.g., a head mask)
(C) a heated, chemically inert gas sample line leading from the flow transducer to the mass spectrometer
(D) a medical ergometer for imposing certain workloads on the test subject
(E) the PTR-MS for measurement of volatile compounds in exhaled breath.

A schematic diagram of our setup is given in Fig. 1 and will be described in detail in the following.

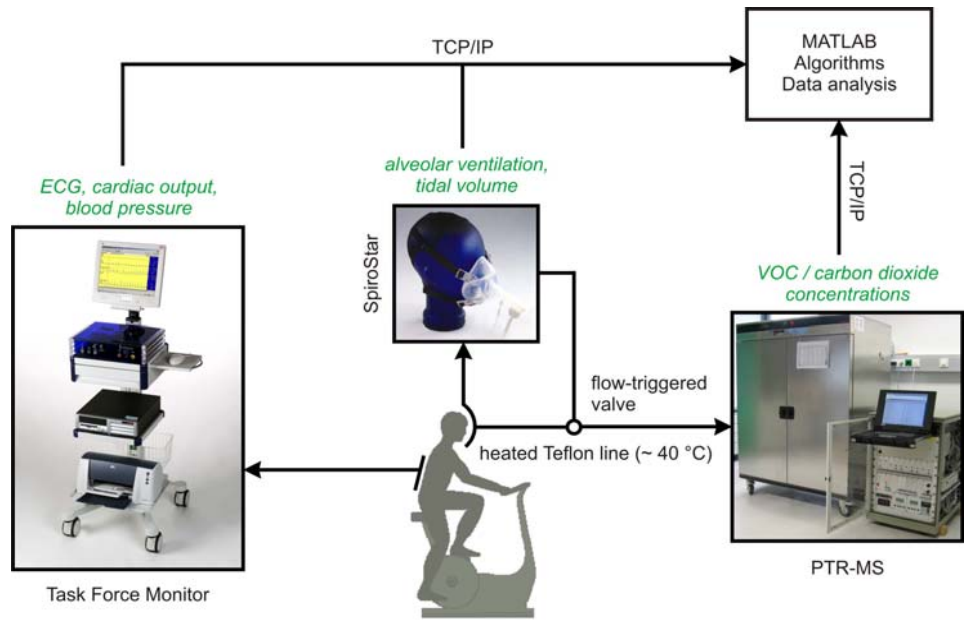

*Figure 1:* Sketch of experimental setup; italic items correspond to accessible variables

With regard to (A), a non-invasive hemodynamic analysis system (Task Force Monitor (TFM), CNSystems, Graz, Austria) was used to determine hemodynamic variables on the basis of standard ECG leads and transthoracic impedance cardiography (ICG). In particular, the device allows the continuous recording of blood pressure and cardiac output with beat-to-beat resolution, which is an essential requirement for relating changes in VOC concentrations to quick hemodynamic variations. This real-time capability ranges among the primary reasons for preferring impedance cardiography over other existing approaches for measurement of heart minute volume as discussed below. Since the user interface of the TFM does not export the corresponding parameter values sequentially, we included a simple data server, sending the latest available data vector to a specified TCP/IP port every second. ICG methodology for determination of cardiac output has been validated against the invasive gold standard method thermal dilution in a variety of situations [21-23], however, a comparison with healthy test subjects during workload conditions is still lacking. On the other hand, the bioimpedance approach has shown good agreement with other non-invasive techniques for obtaining heart minute volume at different exercise levels [24].

Respiratory flow is obtained by means of a Medikro SpiroStar USB differential pressure sensor (Medikro Oy, Kuopio, Finland). Inhalation and exhalation occurs through pre-calibrated, single-use flow transducer mouthpieces, which can be connected to sterilizable or disposable silicone head masks (Cortex Biophysik GmbH, Leipzig, Germany). This enables the test subject to breathe freely through mouth and/or nose while simultaneously reducing the risk of hyperventilation. Although some condensation could occur within the mask, this was not considered a critical issue: gas samples are drawn from the axial mainstream through Luer lock connecting sockets situated about 2 cm from the lips, so possible loss of highly soluble VOCs can be neglected due to the high flow. Based on the Bernoulli principle stating that the constriction of a given airflow will cause a difference in pressure which subsequently can be used to determine the associated flow [25], the device delivers volumetric flow rates ($vf$ [ml$_{BTPS}$/s], i.e., corrected for body temperature and saturation with water vapor) within the flow transducer with a sampling frequency of 100 Hz. As will be illustrated below, proper integration of this signal allows the efficient extraction of actual tidal volume and breathing frequency. The Medikro SpiroStar package comes with a DLL driver implementing functions for immediate communication with the spirometer hardware (initialization, data reading, etc.), which can easily be incorporated into any C based development environment.

Gas sampling is accomplished by a 3 m long, 1/4´´ Teflon tube which is heated using an isolated heating wire (TNI Medical, Freiburg, Germany). The heating is necessary to avoid condensation of water vapor from exhaled breath within the sample line: condensed water droplets would attract hydrophilic compounds, thus depleting the gas sample and leading to erroneous measurement results. More specifically, care was taken to ensure that temperature is kept well over 40 °C along the entire length, thereby guaranteeing that no water condensation occurs. The gas sample line can be connected to the spirometer flow transducer by a metal Luer lock.

Exercise tests are carried out on a computer-controlled, supine medical ergometer (eBike L, GE Medical Systems, Milwaukee, USA) operating at constant levels of power independently of the pedal speed. The supporting bed stabilizes the torso of the volunteer thereby reducing movement artifacts appearing in the acquired physiological signals. The longitudinal tilt can be adjusted from 45° during normal operation to 0° (supine position).

*PTR-MS setup:*

Here we give a short description of the high-sensitivity PTR-MS used in our laboratory (Ionicon Analytic GmbH, Innsbruck, Austria; three turbo pumps), see Fig. 2. The breath source (1), e.g., a Tedlar bag or a volunteer breathing into an *real-time* sampling system as described above is connected to a heated Teflon bypass used to direct the sample gas from the breath source to the outside air by means of a vacuum pump (2). The corresponding flow can be adjusted using a needle valve (3). Assuming ambient pressure conditions in (1), a pressure regulator implemented along the bypass keeps pressure levels at the branching point (4) at a constant value of 0.655 bar, thereby guaranteeing stable pressure conditions of approximately 2.3 mbar in the drift chamber of the PTR-MS (5), which is connected to (4) via a 1/16´´ capillary heated up to 50 °C.

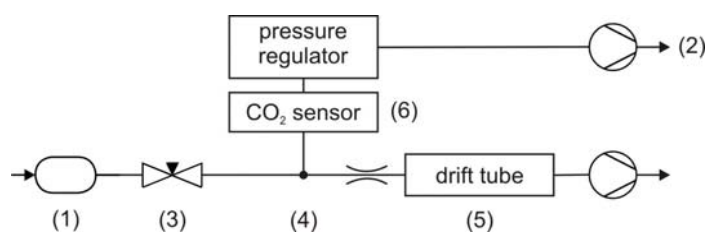

*Figure 2:* Inlet-flow architecture of the PTR-MS; (1) breath source, (2) vacuum pump, (3) needle valve, (4) pressure control point (0.655 bar), (5) drift tube, (6) $CO_2$ sensor

As has been described extensively elsewhere [3, 4], within the drift chamber compounds with higher proton affinities than water are ionized by reacting with hydronium ions ($H_3O^+$) originating from a hollow cathode ion source adjoining the drift tube. The underlying reaction here is

$$H_3O^+ + M \rightarrow MH^+ + H_2O,$$

i.e., reactant gas particles M are protonated to give product ions $MH^+$, which are then separated according to their distinctive mass-to-charge ratios (m/z) by a quadrupole mass spectrometer. Finally, an ion detection system measures count rates $i(H_3O^+)$ and $i(MH^+)$, which can be converted to concentration levels (parts per billion (ppb) – parts of the compound in $10^9$ parts of air) of the compound in question by taking into account substance-specific reaction rates as well as possible fragmentation patterns [26, 27]. The PTR-MS control software provides the count rates associated with the mass-to-charge ratios under study as well as the reaction conditions, i.e, drift voltage (600 V), pressure (2.3 mbar), and temperature (52 °C) within the drift chamber. Again all available data are sent to a predefined TCP/IP port for subsequent storage. In our experiments we limited ourselves to the following six mass-to-charge ratios (corresponding dwell times are given in brackets):

- m/z 21 ($^{18}$O isotope of the hydronium primary ions $H_3O^+$) [500 ms]
- m/z 37 (protonated water dimer, a precursor ion) [2 ms]
- m/z 32 ($O_2^+$, a parasitic precursor ion, adjusted to be below 2% of $H_3O^+$) [10 ms]
- m/z 18 ($NH_4^+$ which – when produced in the hollow cathode – acts as a parasitic precursor ion, adjusted to be below 2% of $H_3O^+$) [10 ms]
- m/z 59 (protonated acetone) [200 ms]
- m/z 69 (protonated isoprene) [200 ms],

resulting in a total cycle duration of about 1.5 seconds. The first two mass-to-charge ratios are necessary for the correct quantification of acetone and isoprene, which we implicitly assume to be the only VOCs contributing to the associated m/z signal [27, 28]. Calibration for these two compounds with different levels of water vapor content was carried out either manually (in the case of isoprene) or using a Gaslab gas mixing unit (Breitfuss Messtechnik GmbH, Harpstedt, Germany) [27]. In particular, the presented concentration levels are determined with respect to the usual water vapor content in exhaled breath of about 6%. Slight variations of water content in the drift tube due to fluctuations in the amount of water clusters originating from the source were assumed to be negligible. For further details regarding calibration factors and fragmentation issues we refer to the "most abundant" approach as described in [29]: the corresponding calibration coefficients for isoprene and acetone were 0.95 and 0.79, respectively, calculated using the standard reaction constant $2.0 \times 10^{-9}$ cm$^3$/s and an E/N ratio of ~ 126 Td.

Due to its low proton affinity, carbon dioxide cannot be measured by PTR-MS. As will be discussed in the following section, $CO_2$ content serves as a well understood control value [30] for assessing the extraction quality of breath gas samples. We therefore included an infrared $CO_2$ sensor (AirSense Model 400, Digital Control Systems, Portland, USA) in our setup (cf. Fig. 2 (6)). Particularly, by placing the sensor behind the pressure controlled branching point, we are able to determine concentrations independently of ambient pressure. Calibration was done with test gas consisting of 5% $CO_2$ in synthetic air (Linde Gas GmbH, Stadl-Paura, Austria). Due to limited flow through the bypass, the $CO_2$ sensor in the PTR-MS acts as a mechanical low-pass filter, with a measurement-induced delay of approximately 45 seconds. Current $CO_2$ concentration is appended to the PTR-MS data vector after each measurement cycle.

*Standardized real-time breath sampling:*

Common measurement practice in breath gas analysis tacitly makes use of the Farhi equation (c.f. Eq. 1, later in this text), which states that – assuming constant respiratory and hemodynamic flow, e.g., during resting conditions – alveolar concentrations of blood borne endogenous VOCs are proportional to their respective concentrations in mixed venous blood and therefore – by taking into account partition coefficients – to tissue levels. This makes alveolar concentration the decisive value in the quantification of blood borne volatile species. However, the extraction of pure alveolar air is hampered by several obstacles, the major ones being mixing with (fresh) anatomic dead space volume as well as exchange in the conducting airways. The latter effect mainly relates to highly soluble substances interacting with mucus linings and will be discussed in conjunction with acetone measurements in the experimental section. In contrast, VOCs with low blood solubility (e.g., isoprene with a Henry constant of ~ 0.029 M/atm [2, 31]) originate almost exclusively from the alveolar blood-gas exchange, resulting in drastically reduced dead space concentrations. Alveolar levels of such molecular species are thus best reflected by end-tidal concentrations, i.e., it is recommendable to discard the first portion of exhaled breath in the analysis process. In the present framework this suggests that any real-time extraction procedure should allow breath to be conducted to the PTR-MS inlet only during the end-tidal fraction of each exhalation phase. In particular, this strategy actively prevents room air from being physically sampled during inhalation, thereby simplifying data evaluation (no exhalation tracking is required).

One possible realization is the implementation of an automatic shutter or valve along the gas sample line, cutting off the connection between mass spectrometer and flow transducer during other time periods (cf. Fig. 1). During valve closing times the sample line then must serve as a sufficiently large buffer volume in order to eliminate potential problems induced by substantial pressure fluctuations within the drift chamber due to continuous bypass deflation by the vacuum pump (cf. Fig. 2). For a fixed sample line length, this amounts to manually adjusting the bypass flow in such a way that two factors are balanced:

    a) bypass flow should be high enough to minimize transport time from flow transducer to drift chamber (i.e., to minimize analysis delay)

    b) bypass flow should be low enough to avoid running into bypass vacuum during valve closing times

Particularly, in our case empirical optimization leads to an analysis delay of approximately 10 seconds. Consequently, about 3 subsequent end-tidal phases are mixed in the sample line during normal breathing. Similarly to $CO_2$ data, the sample line thus represents a mechanical low-pass filter for the count rates delivered by PTR-MS, leading to a slight smoothing of the observed signal. The aforementioned standard setup represents a good trade-off for most test subjects, limiting excessive drift chamber pressure drops to a few isolated cases per measurement sequence. Persons exhibiting very shallow, sustained breathing patterns can be covered by further reducing bypass flow, which however leads to a prolongation of analysis delay.

*Software algorithms:*

In the following, a C++ interface PROCESS_FLOW for consistent on-line shutter control and calculation of respiratory variables is presented. The interface continuously provides tidal volumes, alveolar ventilation and valve opening/closing times on the basis of the signal obtained via the spirometer hardware driver (see above). More specifically, the volumetric flow rates $vf(t)$ are processed sequentially according to two basic heuristics, as shown in Fig. 3. First, we want to neglect small fluctuations due to movement of the head mask or the spirometer's pressure tube. Valid inhalations/exhalations are therefore detected by the first time instant $t^*$ after a zero crossing such that $vf(t^*)$ is smaller/greater than a user selected threshold and the integral from the last baseline crossing to $t^*$ (i.e., the total volume inhaled/exhaled so far) is greater than a predefined dead space volume (representing anatomical dead space and flow transducer volume, i.e., approximately 200 $ml_{BTPS}$ [32]). Subsequently, the last zero crossing stored is accepted as starting point for the current inspiration/expiration phase. Second, if two consecutive inhalations/exhalations occur without at least one valid exhalation/inhalation in between (being the most common non-regular respiratory maneuver during normal breathing), these multiple inhalations/exhalations are treated as one single inhalation/exhalation.

As soon as a new volumetric flow rate is provided, it has to be decided whether the valve should be open or closed. A useful rule here is to consider a certain modifiable percentage (say 50%) of the median of the preceding 3 or 5 exhalation times. This is motivated by the following observation: during regular breathing, after half of the total exhalation time has passed it seems safe to assume that only end-tidal air is being exhaled. If the breathing pattern does not change, a viable shutter regime opens the valve if the following three flow-related conditions are fulfilled:

1. a valid exhalation phase has been identified
2. 50% of the last exhalation time has passed since the starting point
3. $vf(t)$ is greater than a user selected minimal flow $vf_{min}$

The last item accounts for slightly delayed electronic valve response, so we can guarantee actual valve closing to be completed before the onset of inhalation. As a slight generalization of the second requirement we will consider a function of the preceding *n* exhalation times rather than only the last single value: it is well known that the median of *n* values is a robust average estimator, which discards outliers in the data. Consequently, computing the median of the last few exhalation times will filter out extremely short or long exhalation phases caused by coughing, etc. thus maintaining a proper valve control after such breathing events. Fig. 3 summarizes the aforementioned features of the algorithm. A commented version of the C++ interface is available under http://realtime.voc-research.at.

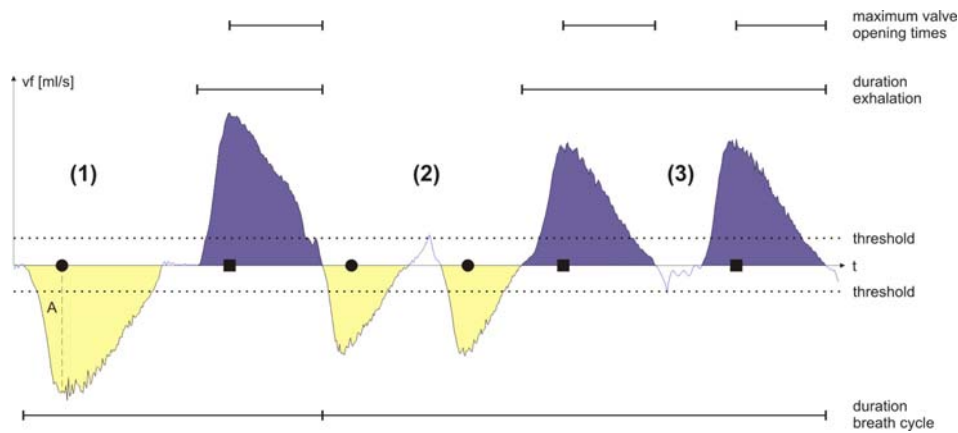

*Figure 3:* Example of the spirometer algorithm

During Phase (1) the flow sensor delivers negative volumetric flow rates indicating the onset of an inhalation *phase*. However, inhalation is only detected if $vf$ is below a predefined threshold and if the inhaled volume (A) exceeds the specified dead space volume $V_D \sim 200$ ml$_{BTPS}$. The filled circle in Fig. 3 marks the first time instant $t^*$ after a zero crossing where these two requirements are fulfilled. The next zero crossing terminates the actual inhalation *segment*. Exhalation is identified analogously, indicated by the filled square. The corresponding time instant guarantees the end of the preceding inhalation *phase* and is associated with two events:

a) the integral (= area under the curve $vf(t)$) calculated during the last inhalation *phase* is accepted as total inhalation volume $V_T$

b) we have arrived at the first possible valve opening time.

However, as explained above, the shutter will usually still remain closed until a user defined percentage of the median of the last preceding exhalation times has passed since the start of exhalation. This requirement can be seen as an additional precaution to avoid the dilution of breath samples in the presence of inter-individually varying dead space air. Then the valve will be open at each time instant where $vf$ is greater than $vf_{min}$ (~ 20 ml/s, not indicated in Fig. 3). As soon as the next zero crossing is encountered, the shutter will be closed until the three requirements stated above are fulfilled again. The detection of a valid inhalation after a valid exhalation in Phase (2) completes the previous breath cycle and is characterized by the following updates:

(a) determination of the preceding exhalation time

(b) determination of the duration of the preceding breathing cycle

(c) calculation of the current alveolar ventilation $\dot{V}_A$ [ml$_{BTPS}$/min], i.e., the flow effectively taking part in pulmonary gas exchange by the usual formula [30]

$$\dot{V}_A = f(V_T - V_D)$$

where the breathing frequency $f$ can be extracted from (b). Thereafter, $V_T$ as well as $\dot{V}_A$ are sent to a specified TCP/IP port. The current breathing frequency can be recovered from these two values by inverting the formula given above. The algorithm moreover accounts for possible multiple inhalation/exhalation phases as sketched in Phase (2) and Phase (3). Accordingly, if two inhalation segments are identified without a valid exhalation in between, the two corresponding inhalation volumes are added to give $V_T$ at the detection of the next exhalation. Similarly, in Phase (3), two expiration segments occur without being separated by a valid inhalation. Hence, the corresponding exhalation times will be merged in further calculations. However, valve control still applies separately to each exhalation phase.

We now have at hand a real-time valve control algorithm guaranteeing the sampling of end-tidal air during the entire measurement sequence while simultaneously computing inhalation volumes, breathing frequency and alveolar ventilation. The shutter is implemented by a Teflon valve (Parker, Fairfield, USA), heated to avoid condensation and placed near the mouth to minimize dead space volume. Opening and closing is accomplished by a standard serial interface realizing the actual status information at each time instant provided by the C++ interface PROCESS_FLOW. Preliminary validation of the presented sampling scheme was done by comparing the $CO_2$ content delivered by the additional sensor in our previously described PTR-MS setup with independent continuous $CO_2$ data simultaneously acquired by means of an IRMA infrared probe (PHASEIN AB, Danderyd, Sweden), cf. Fig. 4. Levels clearly correspond to end-tidal phases, reconfirming the capability of the sampling regime to exclusively extract the last segment of exhaled breath. Alternatively, automatic identification of an end-expiratory phase as well as associated sampling procedures could also be based on any respiratory signal approximately proportional to the aforementioned flow rates. Natural candidates here are $CO_2$ fraction (by defining a threshold detecting alveolar air), temperature (thermistor sensors) for sleep laboratory applications or even PTR-MS signals themselves as suggested in [18]. In any case the limiting factors will be patient convenience, possible signal delay and a sampling frequency which necessarily is much higher than the normal breathing frequency, so the adequacy of underlying method strongly depends on the experimental setup considered. Preliminary comparisons between a hypothesized control algorithm taking $CO_2$ data with breath-to-breath resolution and the presented flow based scheme suggest a good agreement between the two methods.

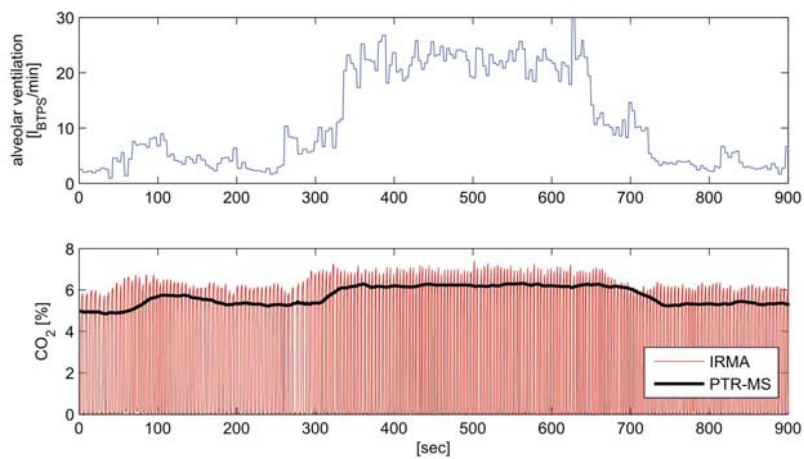

*Figure 4:* Comparison between *real-time* (time resolution ~ 60 ms) and PTR-MS $CO_2$ data during a 75 W ergometer challenge starting at 300 s and ending at 650 s; the measurement-induced delay of the PTR-MS $CO_2$ sensor of about 45 s is readily discernible

*Data acquisition:*

Regarding data acquisition, a simple recording tool (RETBAT – REal Time Breath Analysis Tool) was implemented as a MATLAB graphical user interface. The software can be compiled to be used as a standalone executable without installation of the MATLAB main application. It asynchronously reads in PTR-MS-, spirometer- and TFM-related data from the aforementioned TCP/IP connections and displays them in real-time. Synchronization is done on the basis of available TFM data, which is updated every second. More specifically, as soon as RETBAT receives the actual TFM data vector it accepts the latest available spirometer and PTR-MS data points as current values. The reason for letting the Task Force Monitor application act as a timer is that we want the data collection to run in parallel with the specified ergometer workload scenario which is set by a control application running on the Task Force Monitor PC. Thus, sampling with RETBAT is fully independent of local system time, i.e., the software can be executed on any network-authorized laboratory PC without prior time synchronization.

Before starting the workload scenario, the user has to provide measurement conditions like room temperature, ambient pressure and room air (background) count rates of the mass-to-charge ratios under study. The latter could be used for corrective purposes, however generally the levels are low enough to be neglected in case of isoprene and acetone [27, 28]. After a plausibility check of the provided values, the received physiological data are plotted sequentially and can be explored in real-time by means of the MATLAB plotting tools. Particularly, the count rates of m/z 59 (acetone) and m/z 69 (isoprene) are immediately converted to ppb on the basis of measured drift tube temperature and pressure as described previously. Finally, recorded data will be corrected with respect to the time response delays discussed before and saved to a predefined MATLAB structure which can further be exported to EXCEL. In particular, count rates and ppb levels corresponding to time instants where drift chamber pressure dropped below 2.2 mbar due to excessive shutter closing times (see above) are considered as missing values. This threshold of 2.2 mbar was chosen on the basis of a maximum tolerable variability of room air concentrations during artificially induced pressure fluctuations.

*Experiments:*

We concentrated our efforts on two compounds which have received wide attention in the field of exhaled breath analysis: isoprene and acetone. The reason for this is twofold. First, we wanted to generate comparable data sets in order to reliably validate our sampling protocol. Second, due to their contrasting physical-chemical properties (isoprene is strongly lipophilic whereas acetone is hydrophilic) we view these two species as paradigmatic examples revealing valuable information on the broad spectrum of possible VOC responses according to distinct physiological conditions. In the following, we will briefly review some of the most important facts that have provided a deeper understanding of the involved mechanisms influencing the behavior of aforementioned compounds in exhaled breath.

*Isoprene:*

2-methyl-1,3-butadiene, better known as isoprene (CAS number 78-79-5), is a colorless liquid organic hydrocarbon with a molar mass of 68.11 g/mol and a boiling point of 34 °C. Usually obtained from petroleum and coal to make synthetic rubber it is also the major hydrocarbon which is endogenously produced by mammals [33]. Its primary source is attributed to the mevalonate pathway of cholesterol biosynthesis [34, 35]. Originating from acetyl-CoA mevalonate is transformed into dimethylallyl pyrophosphate from which isoprene is produced. In exhaled human breath isoprene concentration exhibits a large variability. Typical levels in adults during rest have been reported to spread around 100 ppb [36], whereas children show lower levels [37, 38]. Corresponding blood concentrations have been shown to vary around 37 nmol/l [39] with an associated endogenous production rate ranging from 0.15 to 0.34 $\mu$mol/h/kg body weight [31, 40]. In the first reference, metabolization has been quantified as 0.31 $\mu$mol/h/kg body weight, which leads to a net isoprene production of 0.03 $\mu$mol/h/kg body weight (amounting to approximately $5\times10^{-5}$ mol/day for a 70 kg person). Here net isoprene production is defined to be the endogenous production minus metabolization. Body tissue represents a potential storage volume for isoprene in the human body and this is in particular so for fat tissue as can be deduced from the high fat:blood partition coefficient of ~ 82 [31].

Its high abundance in human breath and the fact that there are no indications for concentration changes due to food uptake [41] or for production and release in the upper airways [42] makes isoprene a relatively easily quantifiable test compound. Apart from being a convenient choice in terms of measurability, breath isoprene has been suggested as a sensitive indicator for assaying several metabolic effects in the human body [43]. First, being a possible by-product of cholesterol biosynthesis, it might serve as an additional diagnostic parameter in the care of patients suffering from lipid metabolism disorders such as hypercholesterolemia, which is an established risk factor for atherosclerosis and coronary heart disease. As an example, the estimation of endogenous isoprene production rates on the basis of available breath concentrations as well as appropriate kinetic models might be an adequate tool to determine the contribution of endogenous cholesterol release to the overall serum cholesterol level, thereby improving the diagnostic potential of standard blood tests, which merely quantify the combined effects of endogenous and dietary factors. Moreover, evidence points towards a strong linkage of breath isoprene levels to different physiological states, thus promoting its general use in bio-monitoring, e.g., during sleep [12, 44]. Due to its low blood solubility (Henry's law constant) and boiling point, it is reasonable to assume that exhaled breath concentrations are substantially affected by alveolar ventilation and perfusion (i.e., alveolar minute flow and cardiac output). Significant correlations between cardiac output and breath isoprene concentrations during cardiovascular surgery can be expected. Drastically increasing levels of isoprene concentration were reported [2, 36] at the onset of physical exercise. These findings indicate that breath isoprene measurements might provide new tools for continuous, non-invasive monitoring of cardiac output.

*Acetone:*

Acetone (CAS number 67-64-1), also known as propanone, has a molar mass of 58.08 g/mol. It is one of the ketone bodies, together with beta-hydroxybutyric acid and acetoacetic acid. Acetone is a product of the conversion of acetoacetate by elimination of $CO_2$ [45, 46]

$$CH_3COCH_2COO^- + H^+ \rightarrow CH_3COCH_2COOH \rightarrow CH_3COCH_3 + CO_2.$$

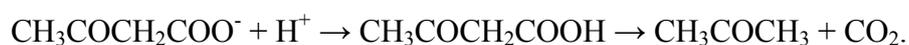

This conversion is either a result of the non-enzymatic decarboxylation of acetoacetate or is catalyzed by acetoacetate decarboxylase. The acetoacetate decarboxylase is induced by starvation and inhibited by acetone itself. High concentrations of blood acetoacetate trigger the acetoacetate decarboxylase, thus draining $H^+$, while acetone, acting as a competitive inhibitor, helps to prevent early acetoacetate decarboxylation of acetoacetate. Acetoacetate is the product of beta-hydroxybutyric acid (= HMG-CoA, an intermediate of the mevalonate pathway) and can either be converted to acetone (see above reaction) or to D-beta-hydroxybutyrate. Acetone is one of the most abundant compounds in human breath. Typical adult exhaled breath concentrations are spread around 600 ppb [27, 47] and plasma concentrations have been quantified as ~ 15 $\mu$mol/l [46]. Moreover, a linear relationship between breath and blood concentrations can be assumed [48]. Blood:tissue solubility was estimated to be 1.38 [49, 50], which makes body tissue a much less efficient buffer for acetone than for isoprene. Because acetone is poorly metabolized [51], simple diffusion and volatilization in the lungs is likely to be the predominant path of removal [48]. Due to its high water-solubility the upper airways however cannot be regarded an inert tube as in the case for isoprene. In fact, the nasal epithelium and as well as the tracheal mucosa linings have been demonstrated to play a critical role in pre-alveolar exchange, a phenomenon which has become known as the wash-in/wash-out effect [50, 52]. More specifically, studies orchestrated in the framework of nasal dosimetry research suggest that up to 75% of the compound inhaled via an exposure chamber is absorbed into the mucous membrane before reaching the alveolar region and almost the entire amount absorbed is released back into the breath stream upon exhalation [51].

Being a byproduct of lipolysis, acetone has often been suggested as a marker compound for monitoring the ketotic state of an individual. Elevated breath acetone levels resulting from fasting are quickly lowered by feeding (as the body is nourished by glucose again [41]) and appear to be correlated with rates of fat loss [53]. No influences of sex, age and BMI on breath concentrations of acetone in adults could be determined [27]. Senthilmohan et al. [1] report slightly increasing values upon physical exercise which again can be rationalized by viewing acetone as metabolite of fat catabolism. Moreover, patients suffering from (uncontrolled) diabetes mellitus have been found to exhibit disproportionately high breath acetone concentrations [54], thus establishing the potential clinical relevance of breath acetone in related medical treatment.

*Test subjects and protocols:*

For our study, 5 males and 3 females with an age range of 25-30 years were recruited as volunteers and agreed to participate in up to three stress ergometer challenges with different workload sequences. The test subjects had to be in good health and physical shape although fitness levels differed. Explicitly non-smokers were chosen even though recent findings did not suggest any difference in isoprene and acetone breath concentration between smokers and non-smokers [55]. Measurements were all done in the morning at approximately the same time when volunteers were able to come into our test laboratory with an empty stomach so that at least 7 h had passed since their last meal. The only exception was drinking of water. Furthermore, volunteers were not allowed to brush their teeth with toothpaste in the morning so we could exclude traces of it as a source of measurement error. No test subject reported any prescribed medication or drug intake. The study was approved by the Ethics Commission of Innsbruck Medical University.

On the day of the experiment the volunteers had to avoid strong physical activity and physical stress on the way from their home to the test laboratory. Following arrival and prior to starting the measurement regime, all test subjects needed to rest for at least 10 min in which they were given instructions regarding the workload protocol. Attention was paid to adjust the test equipment to individual weights and heights and to establish a comfortable seating position during the experiment. Next the volunteers were set up with the Task Force Monitor electrodes, five to obtain the 4-channel ECG and three more for the ICG. Before each measurement, the gas sample line was flushed with nitrogen (purity 6.0, Linde Gas GmbH, Stadl-Paura, Austria) for about one minute. In order to avoid leakage, the head mask was firmly fixed on the volunteer's head by means of a hair net however none of the test subjects reported any discomfort or problems regarding difficulty in breathing or even hyperventilation. The laboratory personnel reminded the volunteer before and during the exercise to minimize torso movements and to breathe regularly. Every event occurring during the measurement was recorded in written documentation including time and event description. Also the staff closely monitored real-time results as they were received through the TCP/IP connections from the different instruments in our Matlab graphical user interface RETBAT.

We created a set of three different protocols (c.f. Fig. 5) all starting with an initial 5 min resting phase without workload. Then the volunteers were challenged to pedal at constant speed between 70-80 r/min on the ergometer which was set up for a workload resistance of 75 Watt for the first 15 min in Protocol 1 and 2. While resting time after this workload sequence was only 3 minutes in Protocol 1 it was extended to 12 minutes in Protocol 2. After a second exercising phase of 15 minutes the resting time was then reversed in both protocols. Both regimes end with a 5 minute workload followed by 5 minutes of final resting. Starting with the same initial 5 minutes of resting, in Protocol 3 the volunteer's position was changed from semi-supine to supine position by lowering the ergometer back rest electronically into a horizontal state for the length of 5 minutes. Subsequently the volunteers were put back into the initial position and after 5 minutes started to pedal with a resistance of 50 Watt. Following an escalating-deescalating regime, this resistance was increased to 100 Watt after 5 minutes and then back to 50 Watt after 10 minutes of exercise. Protocol 3 ended with a 10 minute resting phase.

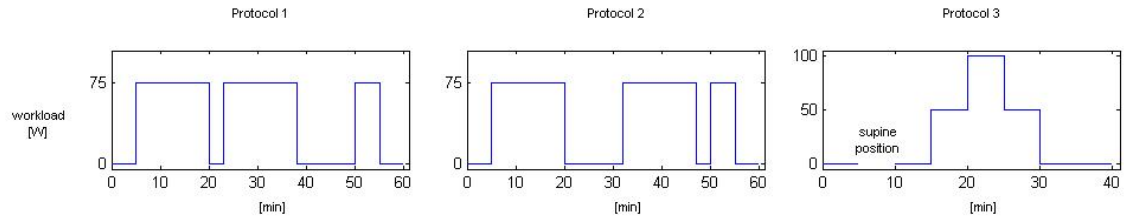

*Figure 5:* Protocols

1. 5 min resting | 15 min exercise (75 W) | 3 min resting | 15 min exercise (75 W) | 12 min resting | 5 min exercise (75 W) | 5 min resting
2. 5 min resting | 15 min exercise (75 W) | 12 min resting | 15 min exercise (75 W) | 3 min resting | 5 min exercise (75 W) | 5 min resting
3. 5 min resting | 5 min supine position | 5 min resting | 5 min exercise (50 W) | 5 min exercise (100 W) | 5 min exercise (50 W) | 10 min resting

The data streams obtained were compiled in a self-contained MATLAB data viewer enabling convenient data exploring and export, which can be downloaded after registration from http://realtime.voc-research.at. Determined physiological variables are summarized in Table 1, together with some nominal values for resting conditions taken from literature. Representative profiles from a single study subject are presented in Fig. 6. Hemodynamic and respiratory variables generally exhibit a very consistent and reproducible behavior among the three protocols. In the following, we will mainly focus on cardiac output and alveolar ventilation. Cardiac output rapidly increased from approx. 5 l/min at rest to a constant plateau of about 12 l/min during permanent workload of 75 W. Simultaneously, alveolar ventilation shows a characteristic rest-to-work transition from 5-10 l/min to a steady state level of approx. 25-30 l/min, thereby increasing the average ventilation-perfusion-ratio by a factor of ~ 3. Transition times from resting conditions to workload steady state and vice versa vary around 5 min. As for the third protocol, changing from semi-supine to supine position usually led to a slight increase of cardiac output while alveolar ventilation remained roughly constant, thereby revealing the individual influence of cardiac output and lung posture on exhaled isoprene and acetone concentrations, respectively.

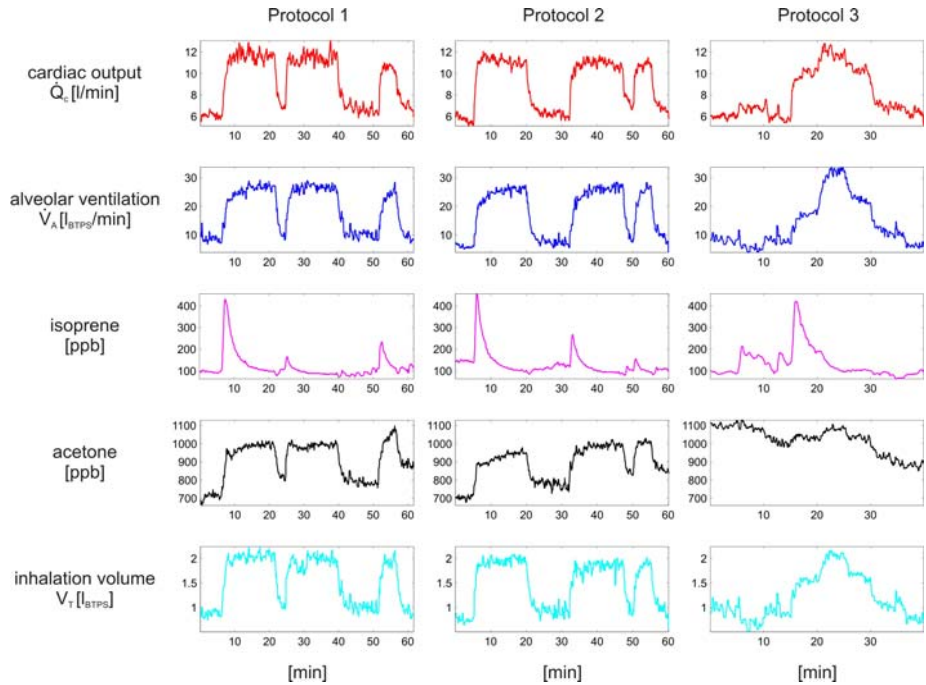

*Figure 6:* Typical results for one single test subject (male, 26 years) according to the three workload scenarios described in the text

## Results

*Results isoprene:*

For all test subjects, end-tidal breath isoprene levels acquired prior to the workload sequence varied around the nominal value of approx. 100 ppb (3.7 nmol/l at body temperature) presented in [28, 36] with minor intra-individual variations, cf. Table 2. Multiplying this level by the measured alveolar minute ventilation leads to a corresponding molar flow (i.e., an amount of isoprene exhaled per minute) of about 30 nmol/min, which is more than 80% of the average net isoprene production for a 70 kg person as discussed above. This indicates that the predominant path of non-metabolic isoprene clearance is via the lungs [56]. Regarding the first two protocols, in accordance with earlier findings [2, 36], the onset of the first exercise period is accompanied by an increase in end-tidal isoprene concentration, usually by a factor of ~ 3-4 within about one minute. Due to a simultaneous increase in ventilation, the associated rise in amount of isoprene exhaled per minute is even more pronounced, leading to a ratio between peak molar flow and molar flow at rest of about 11. This phase is followed by a gradual decline and the development of a new steady state after 15 minutes of pedaling.

Concentrations in this last phase do not differ substantially from the starting values, while molar flow is still higher by a factor of ~ 3 compared to resting levels. In particular, the profile of exhaled isoprene per minute generally is rather different from carbon dioxide output during comparable workload schemes, which typically shows a monotonic rest-to-work transition, cf. [57, 58]. However, one common feature appears to be the abrupt response at the onset of constant workload instituted from rest. The underlying mechanism for this effect remains largely unexplained, but has mainly been ascribed to neurogenic factors affecting ventilation [59].

Interestingly, repeating the same workload procedure described above after intermediate pauses of 3 and 12 minutes, respectively, results in similar concentration profiles but significantly lower peaks, despite almost identical behavior of cardiac output and alveolar ventilation. Consistent effects emerge when reversing the order of the two interceptions, which clearly suggests that initial dynamics tend to be restored with prolonged pauses. For perspective, several follow-up tests indicate that after one hour of rest, maximum values again coincide. There are essentially two hypotheses regarding this effect: (a) changes in mixed venous blood concentration due to depletion/replenishment of an isoprene buffer tissue (e.g., fat), and (b) sustained functional changes in the lung, probably due to recruitment and distension of pulmonary capillaries during exercise [32]. The above-mentioned quantitative considerations and the fact, that breath acetone and carbon dioxide exercise levels (see Fig. 7) do not appear to be affected by preceding pauses [57, 60] favor mechanism (a). However, direct investigation of these issues will have to await future blood tests as in [39].

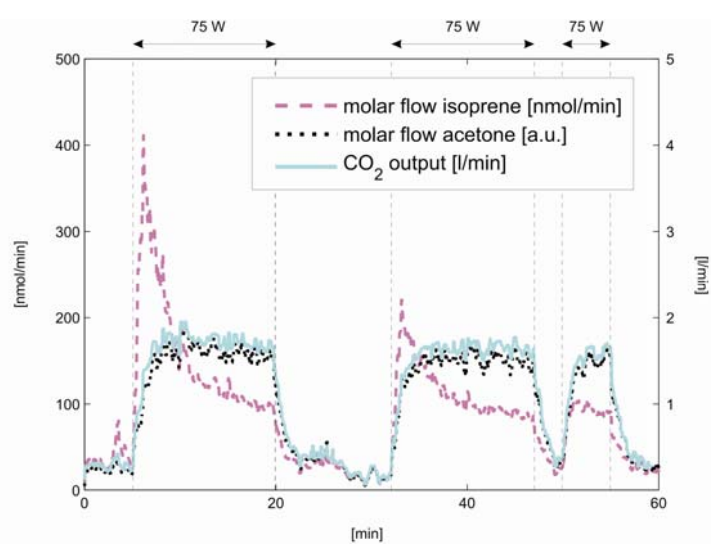

*Figure 7:* Output of isoprene, acetone and $CO_2$ during Protocol 2 (sequential rectangular workload regime of 75 W with intermediate pauses of 12 and 3 minutes, respectively)

Changes of body posture during the third workload scenario generally yield a more or less pronounced rise in breath isoprene concentration, its amplitude being correlated to the associated increase in cardiac output. It should be noted that in such cases, contrary to the behavior during dynamic exercise, isoprene breath concentrations do not appear to revert to baseline levels within a short time. The escalating-deescalating regime performed in the second part reproduces the profile seen in the first two protocols. Specifically, the load step from 50 W to 100 W only has a minor effect on the observed dynamics.

*Results acetone and carbon dioxide:*

Breath acetone concentrations at the beginning of the measurement sequence show typical levels of about 1 ppm. Marked day-to-day variations within one test subject may occur but fall within the range reported in reference [61]. Particularly, as mentioned above, elevated levels might be explained by increased lipolysis due to an empty stomach. Generally, acetone concentrations show higher breath-to-breath fluctuations than isoprene during rest as well as during exercise. The reason for this is still unclear. However, preliminary experiments with our setup indicate that some variation can be attributed to alternate nose and mouth exhalation as well as flow rate. As has already been discussed, due to its high Henry constant, acetone is readily dissolved in the nasal epithelium, leading to lower breath concentrations when the predominant path of exhalation is through the nose [51, 52]. Acetone concentration in exhaled breath during exercise closely resembles the profile of alveolar ventilation respectively inhalation volume, showing abrupt increases respectively drops in the range of 10-40% at the onsets respectively stops of the individual workload periods. Similarly to the results presented by Senthilmohan et al. [1], average concentrations often tend to rise slightly with duration of exercise, which might stem from elevated fat catabolism as a source of energy. Changing to supine position in Protocol 3 seems to have negligible effects. $CO_2$ content initially varies around 4% and exhibits virtually identical dynamics to acetone during the three workload scenarios, with abrupt increase/decrease of ~ 20% at the start/stop of exercise [57]. This is in accordance with Ma et al. [62], who demonstrated a linear correlation between acetone and end-tidal carbon dioxide pressure.

We are aware of the fact, that acetone concentrations obtained with the methodology illustrated above might underestimate alveolar concentrations due to deposition of acetone onto the mucus linings in the conducting airways upon exhalation [49]. In the case of highly water and blood soluble compounds, isothermal rebreathing [63, 64] probably represents the only viable gas sampling scheme to faithfully extract alveolar concentrations. In particular, using this procedure it was demonstrated by Anderson et al. [49] that end-tidal acetone partial pressure is about 20% lower than alveolar partial pressure. However, a straightforward application of isothermal rebreathing in the framework of ergometer challenges has its inherent difficulties, since rebreathing exhalation volumes several times might not be well tolerated during workload segments. Nevertheless, efforts are underway to incorporate this system to our setup. We thereby hope to clarify whether elevated workload breath acetone concentrations observed in our measurements can partly be explained by altered ventilation-perfusion conditions or whether they are simply a result of higher exhalation flow rate and subsequently diminished mucosal absorption as suggested in [49].

**Discussion**

One of the fundamental equations in our present understanding of pulmonary gas exchange is the basic model due to Farhi [65], which expresses mixed alveolar gas concentration $C_A$ of a blood borne gas as a function of its mixed venous concentration $C_{\bar{v}}$, its blood:gas partition coefficient (i.e., dimensionless Henry's law constant) $\lambda$ and an average quotient between alveolar ventilation and capillary perfusion, the ventilation-perfusion-ratio $r = \dot{V}_A / \dot{Q}_c$. Specifically, by describing the lung as one homogenous alveolar unit with an associated end-capillary blood concentration $C_{c'}$ as well as blood inflow/outflow and gas inflow/outflow equal to $\dot{Q}_c$ and $\dot{V}_A$, respectively, conservation of matter leads to

$$V_A \frac{dC_A}{dt} = \dot{V}_A (C_I - C_A) + \dot{Q}_c (C_{\bar{v}} - C_{c'})$$

where $V_A$ is the (invariant) effective pulmonary storage volume of the gas under scrutiny and $C_I$ is its concentration in inspired ambient air, which will usually be close to zero in ordinary isoprene or acetone measurements. Assuming steady state conditions, i.e., neglecting accumulation processes in the lung and requiring that $C_{c'} = \lambda C_A$, i.e., that diffusion equilibrium holds between end-capillary blood and free gas phase (which is a reasonable premise in the case of many VOCs found in exhaled breath [31, 50]), we conclude that

$$\dot{V}_A C_A = \dot{Q}_c (C_{\bar{v}} - \lambda C_A)$$

which can be rearranged to give

$$C_A = \frac{C_{\bar{v}}}{(r + \lambda)} \quad \text{(Eq. 1)}$$

with the alveolar concentration $C_A$ being accessible by exhaled breath measurements. Since left and right heart usually eject the same amount of blood, $\dot{Q}_c$ is commonly set equal to cardiac output. The fact is stressed that the above relation is only valid in the case of physiologically inert gases, but not for oxygen or carbon dioxide, where the purely physically dissolved fraction in blood is small compared to the chemically bound amount. Here $C_{c'} = \lambda C_A$ is replaced by more complicated dissociation functions [66]. Equation 1 of course is a gross simplification of the actual gas exchange conditions within a normal lung, since it completely neglects shunts, physiological dead space and strong regional differences in ventilation-perfusion ratio attributable to gravitational forces and hydrostatic pressure differences [32]. As has been proven in [67], requiring $r$ to be constant throughout the lung corresponds to the implicit assumption of optimal gas exchange and in the case of endogenous VOCs underestimates end-capillary concentrations calculated from alveolar levels.

Nonetheless, the previous relation is one of the pillars for investigating observed behavior of many trace gases found in exhaled breath. First, bearing in mind that during rest on average it holds that $r \sim 1$ [32, 68], we immediately see that breath concentrations of low-soluble trace gases like isoprene ($\lambda_{isoprene} \sim 0.75$ [mol/l / mol/l = dimensionless] at body temperature [31]) are very sensitive to sudden changes in ventilation or perfusion, whereas breath concentrations of compounds with high Henry constants like acetone ($\lambda_{acetone} \sim 200$ [mol/l / mol/l= dimensionless] at body temperature [50, 69]) tend to show a rather damped reaction to such disturbances. Moreover it is evident, that, while other factors are equal, increasing/decreasing alveolar ventilation will decrease/increase exhaled breath concentrations (due to increased/decreased dilution), whereas the relationship between breath concentration and cardiac output is monotonic and reflects dependence on supply. The reader may easily verify that these simple causalities offer a first qualitative explanation for many of the effects observed during the workload scenarios discussed above, particularly isoprene (cf. [2]). However, a precise model elucidating the dynamic characteristics of breath isoprene and acetone concentrations, especially during the unsteady stages of exercise is still lacking. This might be due to the fact that most of the simplifying modeling assumptions allowing for an efficient description of steady state response do not necessarily remain valid in such phases.

One exception is the contribution of Karl et al. [2] who, on the basis of the foregoing deliberations, developed a 2-compartment model in order to reproduce isoprene dynamics in blood and exhaled breath. Their aim was to prove that the large variability of breath isoprene concentration is not due to exercise-induced changes in endogenous synthesis (as for example in the case of flow-dependent release of nitric oxide in endothelial cells [70]), but can mainly be traced back to modified gas exchange behavior. Contradicting the anatomic pulmonary structure, the lung compartment was based on serial instead of parallel arrangement of the alveolar units [32], leading to an exponential rather than rational drop between mixed venous and end-capillary blood concentrations in Equation 1. However, the model response to presented ventilation-perfusion data closely resembled the determined breath concentrations. Unfortunately, this model strongly depends on markedly delayed dynamics of alveolar ventilation compared to cardiac output during exercise, which could not be observed in our measurements: cardiac output and alveolar ventilation increase almost simultaneously [59]. This discrepancy might stem from the fact that alveolar ventilation in [2] was calculated as approximate breathing frequency multiplied by a constant tidal volume, thereby neglecting potential changes in the latter variable, which are also revealed in our experiments. Nevertheless, despite the fact that the model of Karl et al. results in a very poor approximation of breath isoprene concentration given our data, there are several indications that the drastic variation of this value observed during short-term moderate exercise indeed originates from altered gas exchange conditions rather than fluctuations in endogenous production. First, all of the possible biochemical sources of isoprene known up to date are long-term mechanisms, i.e., immediate changes in synthesis rates are not justified by these pathways [34, 71, 72]. Second, taking into account a tissue-lung transport delay of about one minute [73, 74], mixed venous concentration can be assumed constant during the first segment of exercise [58], so possible feedback mechanisms from the body can plausibly be excluded in this period. Third, our data suggest that isoprene breath concentrations can be driven to an elevated plateau by rapidly changing from upright to supine position. This maneuver is very unlikely to induce metabolic variation but rather affects ventilation-perfusion-distribution in the lungs [32].

Accepting the above hypothesis at first glance seems to limit the clinical relevance of breath isoprene, e.g., as a marker compound for therapeutic monitoring of cholesterol related diseases, since well-defined standard (resting) conditions become a fundamental prerequisite for single-breath tests. On the other hand, we are confident that viewing the short-term response of isoprene and other low soluble breath VOCs during workload sequences mainly as lung-induced phenomena can offer entirely novel approaches for the investigation of pulmonary functional properties. This is in line with ongoing efforts to base MIGET (multiple inert gas elimination technique [75, 76]) measurements on endogenous breath compounds rather than intravenously infused inert gases [44], thereby reducing patient load and improving practicability. Here the principal idea is to take advantage of the different solubility and hence distinct exhalation kinetics of several VOCs in order to characterize ventilation-perfusion mismatch throughout the lung, which is of paramount importance in artificial ventilation and serves as a valuable diagnostic tool in the management of patients suffering from pulmonary disorders. Another conceivable application would be (intra-operative) monitoring of cardiac output on the basis of VOC concentrations and ventilation data acquired in real-time.

**Conclusions**

As can be deduced from simple mass balance principles describing pulmonary gas exchange, breath concentrations of blood borne volatile compounds need to be assessed simultaneously with ventilation and perfusion in order to extract comparable and representative values for endogenous levels. Within this framework, an experimental setup efficiently combining PTR-MS measurements with data streams reflecting hemodynamic and respiratory factors was developed, enabling the real-time evaluation of exhaled breath VOC behavior in conjunction with decisive physiological drivers during rest and ergometer-induced workload schemes. Particularly, a methodology for selective breath-by-breath sampling from end-tidal exhalation segments was introduced and validated on the basis of resulting $CO_2$ levels. The key feature of our setup consists of a shutter mechanism separating the PTR-MS from the inhalation/exhalation mouthpiece on the basis of measured respiratory flow. Such an approach has several significant advantages over high-resolution sampling schemes continuously monitoring the entire breath cycle: a larger number of distinct mass-to-charge ratios can be measured, integration times are extended, longer inlet lines are possible and tracking of breath phases is avoided. Moreover, the control algorithm can easily be modified to realize sampling from arbitrary exhalation segments.

In our opinion, pilot studies of breath compound dynamics, e.g., during exercise have to be based on reliably measurable substances, covering prototypic physical-chemical properties.

While isoprene is expected to react very sensitively to changes in ventilation-perfusion ratio due to its low solubility, acetone for analogous reasons shows a comparably stable behavior. Particularly, we were able to reconfirm the experimental findings of Senthilmohan et al. [1] and Karl et al. [2] and added new data which we hope will help to further clarify the kinetics of these species in the human body.

Both acetone and isoprene profiles showed good reproducibility among our moderate workload ergometer stress tests. Data favor the hypothesis that short-term effects visible in the concentration profiles of acetone can be ascribed to different exhalation patterns, while the abrupt response of isoprene at the onset of exercise appears to be caused mainly by changes in pulmonary gas exchange. Some possible clinical applications emerging from this observation have been discussed.

As with every experimental scheme, there are inherent strengths and weaknesses associated with our analysis system: manual fine-tuning of PTR-MS inlet-flow settings is unavoidable for patients exhibiting breathing patterns departing too far from the norm and further optimization is needed in order to reliably guarantee pressure stability within the drift tube. Furthermore, the current setup has a limited applicability in the quantification of highly soluble compounds exchanging in the conducting airways. On the other hand, we are confident that our methodology permits dynamics of non-polar, low-soluble VOCs such as isoprene to be reliably captured over a wide measurement range. Moreover, the suggested sampling algorithm appears general enough to be applicable in other mass spectrometric setups such as SIFT- and IMR-MS as well and hopefully contributes to current standardization efforts in real-time breath sampling.

# Acknowledgements

We are indebted to the referees for numerous helpful suggestions. The research leading to these results has received funding from the European Community's Seventh Framework Programme (FP7/2007-13) under grant agreement no 217967. Julian King is a recipient of a DOC fellowship of the Austrian Academy of Sciences at the Breath Research Unit. Helin Koc gratefully acknowledges financial support by FWF project no Y330. We thank Peter Hamm and Helmut Wiesenhofer for their excellent technical support. We greatly appreciate the generous support of the Member of the Tyrolean regional government Dr Erwin Koler and the Director of the University Clinic of Innsbruck (TILAK) Mag Andreas Steiner.

# References

[1] Senthilmohan, S.T., Milligan, D.B., McEwan, M.J., Freeman, C.G., and Wilson, P.F., *Quantitative analysis of trace gases of breath during exercise using the new SIFT-MS technique.* Redox Rep, 2000. **5**(2-3): 151-3.

[2] Karl, T., Prazeller, P., Mayr, D., Jordan, A., Rieder, J., Fall, R., and Lindinger, W., *Human breath isoprene and its relation to blood cholesterol levels: new measurements and modeling.* J Appl Physiol, 2001. **91**(2): 762-70.

[3] Lindinger, W., Hansel, A., and Jordan, A., *On-line monitoring of volatile organic compounds at pptv levels by means of proton-transfer-reaction mass spectrometry (PTR-MS) - Medical applications, food control and environmental research.* International Journal of Mass Spectrometry, 1998. **173**(3): 191-241.

[4] Lindinger, W., Hansel, A., and Jordan, A., *Proton-transfer-reaction mass spectrometry (PTR-MS): on-line monitoring of volatile organic compunds at pptv levels.* Chem Soc Rev, 1998. **27**: 347 - 354.

[5] Amann, A. and Smith, D., eds. *Breath Analysis for Clinical Diagnosis and Therapeutic Monitoring.* 2005, World Scientific: Singapore.

[6] Hansel, A., Jordan, A., Holzinger, R., Prazeller, P., Vogel, W., and Lindinger, W., *Proton-transfer reaction mass-spectrometry - online trace gas-analysis at the ppb level.* Int J Mass Spectrom Ion Processes, 1995. **149/150**: 609.

[7] Ligor, M., Ligor, T., Bajtarevic, A., Ager, C., Pienz, M., Klieber, M., Denz, H., Fiegl, M., Hilbe, W., Weiss, W., Lukas, P., Jamnig, H., Hackl, M., Buszewski, B., Miekisch, W., Schubert, J., and Amann, A., *Determination of volatile organic compounds appearing in exhaled breath of lung cancer patients by solid phase microextraction and gas chromatography mass spectrometry* Clin Chem Lab Med, 2009. **47**: 550-560.

[8] Ligor, T., Ligor, M., Amann, A., Ager, C., Bachler, M., Dzien, A., and Buszewski, B., *The analysis of healthy volunteers' exhaled breath by the use of solid-phase microextraction and GC-MS.* J Breath Research, 2008. **2**: 046006.

[9] Miekisch, W. and Schubert, J.K., *From highly sophisticated analytical techniques to life-saving diagnostics: Technical developments in breath analysis.* Trends in Analytical Chemistry, 2006. **25**(7): 665-673.

[10] Phillips, M., Altorki, N., Austin, J.H., Cameron, R.B., Cataneo, R.N., Greenberg, J., Kloss, R., Maxfield, R.A., Munawar, M.I., Pass, H.I., Rashid, A., Rom, W.N., and Schmitt, P., *Prediction of lung cancer using volatile biomarkers in breath.* Cancer Biomark, 2007. **3**(2): 95-109.

[11] Phillips, M., Altorki, N., Austin, J.H., Cameron, R.B., Cataneo, R.N., Kloss, R., Maxfield, R.A., Munawar, M.I., Pass, H.I., Rashid, A., Rom, W.N., Schmitt, P., and Wai, J., *Detection of lung cancer using weighted digital analysis of breath biomarkers.* Clin Chim Acta, 2008. **393**(2): 76-84.

[12] Amann, A., Poupart, G., Telser, S., Ledochowski, M., Schmid, A., and Mechtcheriakov, S., *Applications of breath gas analysis in medicine.* Int J Mass Spectrometry, 2004. **239**: 227 - 233.

[13] Amann, A., Spanel, P., and Smith, D., *Breath analysis: the approach towards clinical applications.* Mini reviews in Medicinal Chemistry, 2007. **7**: 115 - 129.

[14] Miekisch, W., Schubert, J.K., Vagts, D.A., and Geiger, K., *Analysis of volatile disease markers in blood.* Clin Chem, 2001. **47**(6): 1053-60.

[15] Rieder, J., Lirk, P., Ebenbichler, C., Gruber, G., Prazeller, P., Lindinger, W., and Amann, A., *Analysis of volatile organic compounds: possible applications in metabolic disorders and cancer screening.* Wien Klin Wochenschr, 2001. **113**(5-6): 181-5.

[16] Miekisch, W., Schubert, J.K., and Noeldge-Schomburg, G.F., *Diagnostic potential of breath analysis--focus on volatile organic compounds.* Clin Chim Acta, 2004. **347**(1-2): 25-39.

[17] Birken, T., Schubert, J., Miekisch, W., and Noldge-Schomburg, G., *A novel visually CO2 controlled alveolar breath sampling technique.* Technol Health Care, 2006. **14**(6): 499-506.

[18] Herbig, J., Titzmann, T., Beauchamp, J., and Kohl, I., *Buffered end-tidal (BET) sampling - a novel method for real-time breath-gas analysis.* Journal of Breath Research, 2008. **2**: 1-9.

[19] Schubert, J.K., Spittler, K.H., Braun, G., Geiger, K., and Guttmann, J., *CO(2)-controlled sampling of alveolar gas in mechanically ventilated patients.* J Appl Physiol, 2001. **90**(2): 486-92.

[20] Fowler, W.S., *Lung function studies; the respiratory dead space.* Am J Physiol, 1948. **154**(3): 405-16.

[21] Fortin, J., Habenbacher, W., Heller, A., Hacker, A., Grüllenberger, R., Innerhofer, J., Passath, H., Wagner, C., Haitchi, G., Flotzinger, D., Pacher, R., and Wach, P., *Non-invasive beat-to-beat cardiac output monitoring by an improved method of transthoracic bioimpedance measurement.* Computers in Biology and Medicine, 2006. **36**: 1185-1203.

[22] Perrino, A.C., Jr., Lippman, A., Ariyan, C., O'Connor, T.Z., and Luther, M., *Intraoperative cardiac output monitoring: comparison of impedance cardiography and thermodilution.* J Cardiothorac Vasc Anesth, 1994. **8**(1): 24-9.

[23] Clancy, T.V., Norman, K., Reynolds, R., Covington, D., and Maxwell, J.G., *Cardiac output measurement in critical care patients: Thoracic Electrical Bioimpedance versus thermodilution.* J Trauma, 1991. **31**(8): 1116-20; discussion 1120-1.

[24] Christensen, T.B., Jensen, B.V., Hjerpe, J., and Kanstrup, I.L., *Cardiac output measured by electric bioimpedance compared with the CO2 rebreathing technique at different exercise levels.* Clin Physiol, 2000. **20**(2): 101-5.

[25] Baker, R.C., *Flow measurement handbook : industrial designs, operating principles, performance, and applications.* 2000, Cambridge, UK ; New York, NY, USA: Cambridge University Press.

[26] Arendacká, B., Schwarz, K., Jr, S.S., Wimmer, G., and Witkovský, V., *Variability issues in determining the concentration of isoprene in human breath by PTR-MS.* Journal of Breath Research, 2008(3): 037007.

[27] Schwarz, K., Pizzini, A., Arendacká, B., Zerlauth, K., Filipiak, W., Dzien, A., Neuner, S., Lechleitner, M., Scholl-Bürgi, S., Miekisch, W., Schubert, J., Unterkofler, K., Witkovský, V., Gastl, G., and Amann, A., *Breath acetone - aspects of normal physiology related to age and gender as determined in a PTR-MS study.* Journal of Breath Research, 2009: accepted for publication.

[28] Kushch, I., Arendacka, B., Stolc, S., Mochalski, P., Filipiak, W., Schwarz, K., Schwentner, L., Schmid, A., Dzien, A., Lechleitner, M., Witkovsky, V., Miekisch, W., Schubert, J., Unterkofler, K., and Amann, A., *Breath isoprene - aspects of normal physiology related to age, gender and cholesterol profile as determined in a proton transfer reaction mass spectrometry study.* Clin Chem Lab Med, 2008. **46**: 1011 - 1018.

[29] Schwarz, K., Filipiak, W., and Amann, A., *Determining concentration patterns of volatile compounds in exhaled breath by PTR-MS.* Journal of Breath Research, 2009: accepted for publication.

[30] Lumb, A., *Nunn's Applied Respiratory Physiology.* 2005, Oxford: Butterworth-Heinemann.

[31]  Filser, J.G., Csanady, G.A., Denk, B., Hartmann, M., Kauffmann, A., Kessler, W., Kreuzer, P.E., Putz, C., Shen, J.H., and Stei, P., *Toxicokinetics of isoprene in rodents and humans.* Toxicology, 1996. **113**(1-3): 278-287.

[32]  West, J.B., *Respiratory Physiology. The Essentials.* 7th ed. 2005, Baltimore: Lippincott Williams & Wilkins.

[33]  Gelmont, D., Stein, R.A., and Mead, J.F., *Isoprene-the main hydrocarbon in human breath.* Biochem Biophys Res Commun, 1981. **99**(4): 1456-60.

[34]  Stone, B.G., Besse, T.J., Duane, W.C., Evans, C.D., and DeMaster, E.G., *Effect of regulating cholesterol biosynthesis on breath isoprene excretion in men.* Lipids, 1993. **28**(8): 705-8.

[35]  Watson, W.P., Cottrell, L., Zhang, D., and Golding, B.T., *Metabolism and molecular toxicology of isoprene.* Chem Biol Interact, 2001. **135-136**: 223-38.

[36]  Turner, C., Spanel, P., and Smith, D., *A longitudinal study of breath isoprene in healthy volunteers using selected ion flow tube mass spectrometry (SIFT-MS).* Physiological Measurement, 2006. **27**(1): 13-22.

[37]  Nelson, N., Lagesson, V., Nosratabadi, A.R., Ludvigsson, J., and Tagesson, C., *Exhaled isoprene and acetone in newborn infants and in children with diabetes mellitus.* Pediatr Res, 1998. **44**(3): 363-7.

[38]  Taucher, J., Hansel, A., Jordan, A., Fall, R., Futrell, J.H., and Lindinger, W., *Detection of isoprene in expired air from human subjects using proton-transfer-reaction mass spectrometry.* Rapid Commun Mass Spectrom, 1997. **11**(11): 1230-4.

[39]  Cailleux, A., Cogny, M., and Allain, P., *Blood lsoprene Concentrations in Humans and in Some Animal Species.* Biochemical Medicine and Metabolic Biology, 1992. **47**: 157-160.

[40]  Taalman, R.D.F.M., *Isoprene: background and issues.* Toxicology, 1996. **113**: 242-246.

[41]  Smith, D., Spanel, P., and Davies, S., *Trace gases in breath of healthy volunteers when fasting and after a protein-calorie meal: a preliminary study.* J Appl Physiol, 1999. **87**(5): 1584-8.

[42]  Lärstad, M.A.E., Torén, K., Bake, B., and Olin, A.-C., *Determination of ethane, pentane and isoprene in exhaled air – effects of breath-holding, flow rate and purified air.* Acta Physiol, 2007. **189**: 87–98.

[43]  Salerno-Kennedy, R. and Cashman, K.D., *Potential applications of breath isoprene as a biomarker in modern medicine: a concise overview.* Wien Klin Wochenschr, 2005. **117**(5-6): 180-6.

[44]  Amann, A., Telser, S., Hofer, L., Schmid, A., and Hinterhuber, H., *Breath gas as a biochemical probe in sleeping individuals*, in *Breath Analysis for Clinical Diagnosis and Therapeutic Monitoring*, Amann, A. and Smith, D., Editors. 2005, World Scientific: Singapore. p. 305 - 316.

[45]  Kalapos, M.P., *Possible physiological roles of acetone metabolism in humans.* Med Hypotheses, 1999. **53**(3): 236-42.

[46]  Kalapos, M.P., *On the mammalian acetone metabolism: from chemistry to clinical implications.* Biochim Biophys Acta, 2003. **1621**(2): 122-39.

[47]  Turner, C., Spanel, P., and Smith, D., *A longitudinal study of ammonia, acetone and propanol in the exhaled breath of 30 subjects using selected ion flow tube mass spectrometry, SIFT-MS.* Physiol Meas, 2006. **27**(4): 321-37.

[48]  Crofford, O.B., Mallard, R.E., Winton, R.E., Rogers, N.L., Jackson, J.C., and Keller, U., *Acetone in breath and blood.* Trans Am Clin Climatol Assoc, 1977. **88**: 128-39.

[49]  Anderson, J.C., Lamm, W.J., and Hlastala, M.P., *Measuring airway exchange of endogenous acetone using a single-exhalation breathing maneuver.* J Appl Physiol, 2006. **100**(3): 880-9.

[50] Mörk, A.K. and Johanson, G., *A human physiological model describing acetone kinetics in blood and breath during various levels of physical exercise.* Toxicol Lett, 2006. **164**(1): 6-15.

[51] Thrall, K.D., Schwartz, R.E., Weitz, K.K., Soelberg, J.J., Foureman, G.L., Prah, J.D., and Timchalk, C., *A real-time method to evaluate the nasal deposition and clearance of acetone in the human volunteer.* Inhal Toxicol, 2003. **15**(6): 523-38.

[52] Anderson, J.C., Babb, A.L., and Hlastala, M.P., *Modeling soluble gas exchange in the airways and alveoli.* Ann Biomed Eng, 2003. **31**(11): 1402-22.

[53] Kundu, S.K., Bruzek, J.A., Nair, R., and Judilla, A.M., *Breath acetone analyzer: diagnostic tool to monitor dietary fat loss.* Clin Chem, 1993. **39**(1): 87-92.

[54] Tassopoulos, C.N., Barnett, D., and Fraser, T.R., *Breath-acetone and blood-sugar measurements in diabetes.* Lancet, 1969. **1**(7609): 1282-6.

[55] Euler, D.E., Davé, S.J., and Guo, H., *Effect of cigarette smoking on pentane excretion in alveolar breath.* Clinical Chemistry, 1996. **42**(2): 303-308.

[56] Dahl, A.R., Birnbaum, L.S., Bond, J.A., Gervasi, P.G., and Henderson, R.F., *The fate of isoprene inhaled by rats: comparison to butadiene.* Toxicol Appl Pharmacol, 1987. **89**(2): 237-48.

[57] Wasserman, D.H. and Whipp, B.J., *Coupling of ventilation to pulmonary gas exchange during nonsteady-state work in men.* J Appl Physiol, 1983. **54**(2): 587-593.

[58] Whipp, B.J., Ward, S.A., Lamarra, N., Davis, J.A., and Wasserman, K., *Parameters of ventilatory and gas exchange dynamics during exercise.* J Appl Physiol, 1982. **52**(6): 1506-13.

[59] Johnson, A.T., *Biomechanics and exercise physiology: quantitative modeling.* 2nd ed. 2007, Boca Raton: CRC Press.

[60] Grassi, B., Marconi, C., Meyer, M., Rieu, M., and Cerretelli, P., *Gas exchange and cardiovascular kinetics with different exercise protocols in heart transplant recipients.* J Appl Physiol, 1997. **82**(6): 1952-62.

[61] Diskin, A.M., Spanel, P., and Smith, D., *Time variation of ammonia, acetone, isoprene and ethanol in breath: a quantitative SIFT-MS study over 30 days.* Physiol Meas, 2003. **24**(1): 107-19.

[62] Ma, W., Liu, X., and Pawliszyn, J., *Analysis of human breath with micro extraction techniques and continuous monitoring of carbon dioxide concentration.* Anal Bioanal Chem, 2006. **385**: 1398-1408.

[63] O'Hara, M.E., O'Hehir, S., Green, S., and Mayhew, C.A., *Development of a protocol to measure volatile organic compounds in human breath: a comparison of rebreathing and on-line single exhalations using proton transfer reaction mass spectrometry.* Physiol Meas, 2008. **29**(3): 309-30.

[64] Ohlsson, J., Ralph, D.D., Mandelkorn, M.A., Babb, A.L., and Hlastala, M.P., *Accurate measurement of blood alcohol concentration with isothermal rebreathing.* J Stud Alcohol, 1990. **51**(1): 6-13.

[65] Farhi, L.E., *Elimination of inert gas by the lung.* Respir Physiol, 1967. **3**(1): 1-11.

[66] Ottesen, J.T., Olufsen, M.S., and Larsen, J.K., *Applied Mathematical Models in Human Physiology*. 2004, Philadelphia: SIAM.

[67] Hoppensteadt, F.C. and Peskin, C.S., *Modeling and Simulation in Medicine and the Life Sciences*. 2nd ed. 2002, New York: Springer.

[68] Csanady, G.A. and Filser, J.G., *The relevance of physical activity for the kinetics of inhaled gaseous substances.* Arch Toxicol, 2001. **74**(11): 663-72.

[69] Kumagai, S. and Matsunaga, I., *A lung model describing uptake of organic solvents and roles of mucosal blood flow and metabolism in the bronchioles.* Inhal Toxicol, 2000. **12**(6): 491-510.

[70] Mohrman, D.E. and Heller, L.J., *Cardiovascluar Physiology*. 6th ed. 2006: Lange Medical Books/McGraw-Hill.
[71] Kuzma, J., Nemecek-Marshall, M., Pollock, W.H., and Fall, R., *Bacteria produce the volatile hydrocarbon isoprene.* Curr Microbiol, 1995. **30**(2): 97-103.
[72] Stein, R.A. and Mead, J.F., *Small hydrocarbons formed by the peroxidation of squalene.* Chem Phys Lipids, 1988. **46**(2): 117-20.
[73] Batzel, J.J., Kappel, F., Schneditz, D., and Tran, H.T., *Cardiovascular and respiratory systems: modeling, analysis and control*. 2007, Philadelphia: SIAM.
[74] Grodins, F.S., Buell, J., and Bart, A.J., *Mathematical analysis and digital simulation of the respiratory control system.* J Appl Physiol, 1967. **22**(2): 260-76.
[75] Wagner, P.D., *The multiple inert gas elimination technique (MIGET).* Intensive Care Med, 2008. **34**(6): 994-1001.
[76] Wagner, P.D., Saltzman, H.A., and West, J.B., *Measurement of Continuous Distributions of Ventilation-Perfusion Ratios - Theory.* Journal of Applied Physiology, 1974. **36**(5): 588-599.
[77] Silbernagl, S. and Despopoulos, A., *Taschenatlas der Physiologie*. 1979, Stuttgart: Georg Thieme Verlag.

# Tables

| Variable | Abbreviation | Nominal value |
|---|---|---|
| **Hemodynamic parameters** | | |
| Heart rate | HR | 70 [bpm] [77] |
| RR interval | RRI | 850 [ms] [77] |
| Systolic blood pressure | sBP | 120 [mmHg] [77] |
| Diastolic blood pressure | dBP | 80 [mmHg] [77] |
| Stroke volume | SV | 70 [ml/min] [77] |
| Cardiac output ($\dot{Q}_c$) | CO | 5 [l/min] [77] |
| Total peripheral resistance | TPR | 1600 [dyne.s/cm$^5$] [77] |
| **Ventilation parameters** | | |
| Alveolar ventilation ($\dot{V}_A$) | ALV | 5.2 [l$_{BTPS}$/min] [32] |
| Inhalation volume ($V_T$) | vinh | 0.5 [l$_{BTPS}$] [32] |
| **PTR-MS related data** | | |
| Drift chamber pressure | pdrift | 2.3 [mbar] |
| Carbon dioxide content (end-expiratory) | pCO2 | 5.6 [%] [30] |
| Count rates m/z * | CR* | |
| Acetone concentration | PPB59 | 500 [ppb] [27, 47] |
| Isoprene concentration | PPB69 | 100 [ppb] [28, 36] |

*Table 1:* Summary of measured variables together with some nominal values during resting conditions

| Subject No. | 1 | 2 | 3 | 4 | 5 | 6 | 7 | 8 |
|---|---|---|---|---|---|---|---|---|
| Protocol | 1 \| 2 \| 3 | 1 \| 2 \| 3 | 1 \| 2 \| 3 | 1 \| 2 \| 3 | 1 \| 2 \| 3 | 1 \| 2 \| 3 | 1 \| 2 \| 3 | 1 \| 2 \| 3 |
| $C_{rest}$ [ppb] | 99 \| 146 \| 86 | 48 \| 44 \| 50 | 75 \| 104 \| 119 | 180 \| 177 \| 126 | 156 \| 108 \| 125 | 64 \| 92 \| 54 | 144 \| 156 \| 128 | 43 \| 46 \| 40 |
| $m_{rest}$ [nmol/min] | 32 \| 34 \| 29 | 22 \| 19 \| 17 | 16 \| 29 \| 29 | 41 \| 43 \| 32 | 40 \| 55 \| 45 | 12 \| 20 \| 8 | 30 \| 34 \| 24 | 9 \| 9 \| 12 |
| $C_{peak}$ [ppb] | 430 \| 456 \| # | 212 \| 142 \| # | 244 \| 243 \| # | 610 \| 656 \| # | 534 \| 322 \| # | 306 \| 376 \| # | 605 \| 495 \| # | 117 \| 203 \| # |
| $m_{peak}$ [nmol/min] | 360 \| 346 \| # | 184 \| 138 \| # | 167 \| 226 \| # | 455 \| 475 \| # | 478 \| 390 \| # | 249 \| 330 \| # | 453 \| 411 \| # | 75 \| 140 \| # |
| $C_{work}$ [ppb] | 101 \| 98 \| # | 58 \| 53 \| # | 52 \| 49 \| # | 169 \| 193 \| # | 130 \| 108 \| # | 45 \| 58 \| # | 92 \| 91 \| # | 30 \| 41 \| # |
| $m_{work}$ [nmol/min] | 108 \| 103 \| # | 60 \| 53 \| # | 50 \| 55 \| # | 167 \| 171 \| # | 136 \| 98 \| # | 60 \| 70 \| # | 94 \| 93 \| # | 32 \| 42 \| # |
| $C_{supine}$ [ppb] | # \| # \| 136 | # \| # \| 56 | # \| # \| 175 | # \| # \| 165 | # \| # \| 126 | # \| # \| 86 | # \| # \| 213 | # \| # \| 45 |
| $m_{supine}$ [nmol/min] | # \| # \| 45 | # \| # \| 28 | # \| # \| 54 | # \| # \| 51 | # \| # \| 50 | # \| # \| 25 | # \| # \| 45 | # \| # \| 15 |
| $C_{peak} / C_{rest}$ | 4.3 \| 3.1 \| # | 4.1 \| 3.2 \| # | 3.3 \| 2.3 \| # | 3.4 \| 3.7 \| # | 3.4 \| 3.0 \| # | 4.8 \| 4.1 \| # | 4.1 \| 3.2 \| # | 2.7 \| 4.1 \| # |
| $C_{work} / C_{rest}$ | 1.0 \| 0.7 \| # | 1.2 \| 1.2 \| # | 0.7 \| 0.5 \| # | 0.9 \| 1.1 \| # | 0.8 \| 1.0 \| # | 0.7 \| 0.6 \| # | 0.6 \| 0.6 \| # | 0.7 \| 0.9 \| # |
| $C_{supine} / C_{rest}$ | # \| # \| 1.6 | # \| # \| 1.1 | # \| # \| 1.5 | # \| # \| 1.3 | # \| # \| 1.0 | # \| # \| 1.6 | # \| # \| 1.7 | # \| # \| 1.7 |
| $m_{peak} / m_{rest}$ | 11.3 \| 10.2 \| # | 8.4 \| 7.3 \| # | 10.4 \| 7.8 \| # | 11.1 \| 11.0 \| # | 12.0 \| 7.1 \| # | 20.8 \| 16.5 \| # | 15.1 \| 12.1 \| # | 8.3 \| 15.6 \| # |
| $m_{work} / m_{rest}$ | 3.4 \| 3.0 \| # | 2.7 \| 2.8 \| # | 3.1 \| 1.9 \| # | 4.1 \| 4.0 \| # | 3.4 \| 1.8 \| # | 5.0 \| 3.5 \| # | 3.1 \| 2.7 \| # | 3.6 \| 4.7 \| # |
| $m_{supine} / m_{rest}$ | # \| # \| 1.6 | # \| # \| 1.6 | # \| # \| 1.9 | # \| # \| 1.6 | # \| # \| 1.1 | # \| # \| 3.1 | # \| # \| 1.9 | # \| # \| 1.3 |

*Table 2:* Summary of observed breath concentrations (C) and molar flows (m) of isoprene for all eight study subjects. Values for the stages rest ($C_{rest}$, $m_{rest}$), work ($C_{work}$, $m_{work}$) and supine position ($C_{supine}$, $m_{supine}$) were obtained by filtering concentration and ventilation raw data by means of a 20 step median filter and calculating the mean values over the following time intervals (where applicable): rest (1-2 min), work (19-20 min), supine position (10-11 min); peak values correspond to maxima of the filtered profiles.